\documentclass[11pt]{article}
\pdfoutput=1
\usepackage{jheppub}
\usepackage{graphicx,float}
\usepackage{amssymb}
\usepackage{subcaption}
\usepackage{amsmath}
\usepackage{slashed}
\usepackage{hyperref}
\usepackage{caption}
\usepackage{xcolor}
\usepackage{dsfont}
\newcommand{\comments}[1]{}

\newcommand\beq{\begin{equation}}
\newcommand\eeq{\end{equation}}

\newcommand{\dd}{\mathrm{d}}

\newcommand\eps{\epsilon}

\DeclareMathOperator{\arccot}{arccot}

\DeclareMathOperator{\arcsinhh}{arcsin(h)}
\DeclareMathOperator{\sinhh}{sin(h)}

\DeclareMathOperator{\arctanh}{arctanh}

 \usepackage{cancel}
\def\shrug{\texttt{\raisebox{0.75em}{\char`\_}\char`\\\char`\_\kern-0.5ex(\kern-0.25ex\raisebox{0.25ex}{\rotatebox{45}{\raisebox{-.75ex}"\kern-1.5ex\rotatebox{-90})}}\kern-0.5ex)\kern-0.5ex\char`\_/\raisebox{0.75em}{\char`\_}}}
\usepackage{bm}
\title{Entanglement entropy and $T\bar T$ deformations beyond antipodal points from holography}
\preprint{\today}

\affiliation[\Diamond]{Theoretisch-Physikalisches Institut, Friedrich-Schiller-Universit\"at Jena,
Max-Wien-Platz 1, D-07743 Jena, Germany.}

\author[\,\Diamond\,,\,\star]{Sebastian Grieninger}

\affiliation[\star]{Department of Physics, University of Washington, Seattle, WA 98195-1560, USA}

\emailAdd{sebastian.grieninger@gmail.com}
\abstract{ We consider the entanglement entropies in dS$_d$ sliced (A)dS$_{d+1}$ in the presence of a hard radial cutoff for $2\le d\le 6$. By considering a one parameter family of analytical solutions, parametrized by their turning point in the bulk $r^\star$, we are able to compute the entanglement entropy for generic intervals on the cutoff slice. It has been proposed that the field theory dual of this scenario is a strongly coupled CFT, deformed by a certain irrelevant deformation -- the so-called $T\bar T$ deformation. Surprisingly, we find that we may write the entanglement entropies formally in the same way as the entanglement entropy for antipodal points on the sphere by introducing an effective radius $R_\text{eff}=R\,\cos(\beta_\epsilon)$, where $R$ is the radius of the sphere and $\beta_\epsilon$ related to the length of the interval. Geometrically, this is equivalent to following the $T\bar T$ trajectory until the generic interval corresponds to antipodal points on the sphere.
Finally, we check our results by comparing the asymptotic behavior (no Dirichlet wall present) with the results of Casini, Huerta and Myers. We then switch on counterterms on the cutoff slice which are important with regards to the field theory calculation. We explicitly compute the contributions of the counterterms to the entanglement entropy by considering the Wald entropy. In the second part of this work, we extend the field theory calculation of the entanglement entropy for antipodal points for a $d$-dimensional field theory in context of DS/dS holography. We find excellent agreement with the results from holography and show, in particular, that the effects of the counterterms in the field theory calculation match the Wald entropy associated with the counterterms on the gravity side.}
\begin{document}
\maketitle
\thispagestyle{empty}

\newpage
\setcounter{page}{1}
\section{Introduction}
One remarkable development in the recent years has been a novel access to irrelevant (non-renormalizable) deformations in two dimensional quantum field theories (QFTs). Unlike the usual irrelevant deformations, the so-called $T\bar T$ deformation~ \cite{Smirnov:2016lqw,Cavaglia:2016oda,Zamolodchikov:2004ce} has the intriguing feature that it is -- unlike the usual irrelevant deformations -- exactly solvable. Starting from a generic seed QFT, we are able to define a trajectory from the IR to the UV in the field theory space triggered by deforming the QFT with a $T\bar T$ deformation in each step. Even through the theory flows towards the UV, we are still able to derive a lot of interesting quantities in exact form simply from possessing an understanding of undeformed theory. These quantities include the finite volume spectrum, the S-matrix and the deformed classical Lagrangian -- all of which have been extensively discussed in the literature ~\cite{ McGough:2016lol,  Dubovsky:2017cnj,Jeong:2019ylz, Cardy:2018sdv, Apolo:2019yfj,Cottrell:2018skz,Guica:2019nzm, Kraus:2018xrn, Donnelly:2018bef, Hartman:2018tkw, Taylor:2018xcy, Bonelli:2018kik,LeFloch:2019rut, Aharony:2018vux,Chakraborty:2019mdf, Gross:2019ach,Aharony:2018bad, Datta:2018thy, Giveon:2017nie,Shyam:2017znq,Murdia:2019fax,Shyam:2018sro,Jiang:2019tcq,Park:2018snf, Guica:2017lia, Baggio:2018rpv, Chang:2018dge,Chen:2019mis,Ota:2019yfe,Cardy:2019qao,Banerjee:2019ewu,Caputa:2019pam,Gorbenko:2018oov,2019arXiv190809299S,Chakraborty:2018kpr} (see \cite{Jiang:2019hxb} for lecture notes). 

An interesting approach to $T\bar T$ deformations is the proposal of a holographic dual by McGough, Mezei, and Verlinde \cite{McGough:2016lol} in order to use the powerful toolkit provided by holographic dualities for studying problems in strongly coupled field theories. From a bulk perspective,
deforming a field theory by an irrelevant deformation has drastic effects on the UV behavior. McGough, Mezei, and Verlinde conjectured to simply chop off the asymptotic region of the spacetime. In other words, deforming the conformal field theory (CFT) by the $T\bar T$ operator is dual to introducing a hard radial cutoff (Dirichlet wall) at a finite radial position $r=r_c$ in the bulk. The hard radial cutoff removes the UV region of the spacetime and the dual field theory which lives on the cutoff surface is no longer conformal. For Anti-de Sitter (AdS) this was more extensively studied in \cite{Kraus:2018xrn}. Note that we are using the AdS/CFT correspondence in the weak form throughout this work which means that we are working with a strongly coupled CFT at large N on the field theory side dual to weakly coupled classical gravity.

One interesting aspect of quantum theories -- especially with regards to quantum information -- is the entanglement of quantum states. The entanglement entropy provides a measure of how much quantum information is stored in a specific quantum state and it may be defined in the universal language of quantum fields (although explicit calculations are extremely difficult to do). Calabrese and Cardy developed a powerful approach to calculate entanglement entropies in QFTs by applying so-called replica trick to entanglement entropy calculations in 2D QFTs~\cite{Calabrese:2004eu}. For strongly coupled field theories, however, there is a very elegant way to compute entanglement entropies. Based on the observation that the Bekenstein-Hawking entropy is proportional to the area of the black hole, Ryu and Takayanagi \cite{Ryu:2006bv} derived that the entanglement entropy of a subsystem may be computed holographically by computing the area of minimal surface in the bulk enclosing the subsystem.  

The authors of \cite{Donnelly:2018bef} were able to give further evidence in favor of the conjecture of \cite{McGough:2016lol} by showing that the entanglement entropy for antipodal points in a two-dimensional CFT deformed by a $T\bar T$ deformation matches the entanglement entropy computed in AdS$_3$ in presence of a hard radial cutoff. This analysis has been extended to higher dimensions \cite{Banerjee:2019ewu,Caputa:2019pam,Hartman:2018tkw,Taylor:2018xcy}, and to dS$_3$ \cite{Gorbenko:2018oov} in the context of the DS/dS duality which we will review shortly. This leads to the question -- what happens to the entanglement entropy for intervals different from antipodal points? On the field theory side, this seems to be a notoriously hard question to ask. The authors of \cite{Chen:2018eqk} were able to calculate the first order corrections for a field theory in Minkowski space while the authors of \cite{Jeong:2019ylz} estimated the entanglement entropy for subintervals. We will answer this question on the gravitational side of the duality and derive the exact form of the entanglement entropy in general dimensions.

While the AdS/CFT-correspondence provides us with a definition of quantum gravity in AdS, quantum gravity in dS has yet to be established. One proposal for how to apply holography to dS is the so-called DS/dS correspondence \cite{Alishahiha:2004md} which is based on uplifting the AdS/CFT correspondence~\cite{Alishahiha:2005dj,Dong:2010pm,Freivogel:2006xu,Dong:2011uf}. The basic idea of DS/dS becomes apparent when we express the metric of $D\!=\!d\!+\!1$-dimensional (Anti-)de-Sitter space with curvature radius $L$ as a warped space given by the metric
\beq
ds^2_{(A)DS_D} = dr^2 + (L \sinhh(r/L))^2 ds^2_{dS_{d}},\label{Adsds}
\eeq 
where the radial direction is denoted by $r$ and the warpfactors $L\,\sin(r/L)$ and $L\, \sinh(r/L)$ correspond to dS and AdS, respectively. In both cases, the warpfactors vanish linearly at the horizon, located at $r/L=0$. In dS, we see that the warpfactor has a maximum at the central ``UV slice" ($r/L= \pi/2$), whereas the AdS warpfactor is growing boundlessly for $r\to\infty$. It is interesting to note that the bulk AdS and dS spacetime are identical in the highly redshifted region $r/L\ll1$ since $\sinhh(r/L)\sim r/L$. For dS$_d$ sliced AdS$_D$ \eqref{Adsds}, we have a well-established description of the CFT living in dS$_d$ in terms of the AdS/CFT-correspondence. Since the two spacetimes are indistinguishable in the IR region, the authors of \cite{Alishahiha:2004md} conjectured that infrared degrees of freedom of the CFT dual to AdS$_D$ are also a holographic dual for the infrared region of dS$_D$. By this identification, we are able to establish a holographic dual to dS. The authors of \cite{Gorbenko:2018oov} showed in $d=2$ how to systematically derive this dual by first starting with the CFT dual to AdS$_D$; by deforming the theory with the $T\bar T$ operator, they were able to remove the UV part of the geometry. In the IR region AdS and dS are identical and CFT dual of AdS (via the AdS/CFT-correspondence) is also the CFT dual of dS; by deforming the theory by yet another $T\bar T$ deformation, we can ``grow back'' the UV part of the spacetime -- this time for DS$_D$ instead of the asymptotic AdS region. One natural question is, how do these $T\bar T$ deformations look in higher dimensions?

The UV regions corresponding to dS and AdS are quite different from one another; the fact that the warpfactor in the dS case reaches its maximum in the UV means that the dual CFT intrinsically possesses a cutoff in the UV. In contrast, the warpfactor of AdS grows without bound. Another difference occurs in dS where there is a second near horizon region beyond the central slice at $r/L=\pi$ -- meaning that there is a second dual CFT. Furthermore, the author of \cite{Karch:2003em} showed that the dual CFT also contains dynamical gravity.

Last but not least, since the origin of the DS/dS duality being the AdS/CFT correspondence, we may infer how to calculate entanglement entropies in the $d$-dimensional field theory in terms of minimal surfaces in the $D$-dimensional geometry \cite{Ryu:2006bv} as was explored in \cite{Dong:2018cuv,Geng:2019bnn}. In fact, the authors of \cite{Geng:2019bnn} found a one parameter family of entangling surfaces which all reproduce the dS entropy correctly. This means that independent of the turning point of the entangling surfaces in the bulk, we will always end up with the same area.

The paper is organized as follows: the first part consists of deriving the entanglement entropies for arbitrary intervals in (A)dS$_D$ in presence of a hard radial cutoff. This extends the results of \cite{Banerjee:2019ewu,Donnelly:2018bef,Gorbenko:2018oov} from antipodal points to generic intervals in both AdS and dS. In the second part, we generalize the work of \cite{Gorbenko:2018oov} to higher dimensions. We compute the entanglement entropies on the field theory side for a CFT deformed by a $T\bar T$ deformation dual to dS$_D$ with a hard radial cutoff. The calculation follows \cite{Banerjee:2019ewu}, where this has been derived for a field theory dual to AdS$_D$ with a hard radial cutoff. Finally, we compare the field theory results to the results obtained from the gravitational theory.  
\section{Dirichlet walls and Entanglement Entropy in holography}\label{EEholography}
In this section, we will compute the entanglement entropy in (A)dS with a Dirichlet wall, that is located at $r=r_c$. We consider the metric \eqref{Adsds} for (A)dS$_D$ in dS slicing in static coordinates\footnote{We consider only one of the two static patches in dS which means we are restricting $r/L$ to $[0,\pi/2]$.}
\begin{equation}
    ds^2=dr^2+(L\,\sinhh(r/L))^2\left(-\cos^2(\beta)\,d\tau^2+d\beta^2+\sin^{2}(\beta)\,d\Omega_{D-3}^2\right).\label{metricAdsdS}
\end{equation}
In these coordinates, the horizon is located at $r=0$, the AdS boundary at $r=\infty$ and the dS central slice at $r/L=\pi/2$.  We want to calculate entanglement entropies associated with spherical entangling surfaces centered around the center of the static patch for an observer located at
\begin{equation}
    \tau=0\quad \quad\quad\beta=\beta_0\in[0,\pi/2].
\end{equation}
According to the Ryu-Takayanagi formula, our task at hand is to calculate the surface minimizing the area. We will do this by committing to a parametrization and determining the entangling surfaces by solving the Euler-Lagrange equations.
\subsection{Dirichlet walls and Entanglement Entropy in dS}
We start by studying the entangling surfaces in dS. This has been done in previous work by the author in \cite{Geng:2019bnn}. Concretely, the authors found a one parameter family of entangling surfaces which all correctly reproduce the dS entropy. These surfaces may be found by considering
 the standard ``U"-shaped surfaces that are hanging down towards the IR and are parametrized in terms of $\beta(r)$\footnote{Throughout this paper, we shifted the radial coordinate $r/L$ by $\pi/2$ compared to \cite{Geng:2019bnn} for pedagogical reasons. This leads to a warpfactor of $\sin(r/L)$ instead of $\cos(r/L)$ and helps us to establish a consistent notation with the AdS case which requires $\sinh(r/L)$ as warpfactor.}
\beq
{\cal L}_{I} = L^{D-3}\, \cos^{D-3}(\beta) \sin^{D-3} \left ( \frac{r}{L} \right ) \sqrt{1 +L^2\, \sin^2 \left ( \frac{r}{L} \right )
 (\beta')^2}.\label{lagr}
\eeq
The equations of motions associated with the Lagrangian are solved by~\cite{Geng:2019bnn}
\beq
\label{analyticsolution}
 \beta(r)=\arcsin\left[\tan\left(r^\star/L\right)/\tan\left(r/L\right)\right],
\eeq
where $r^\star$ is the turning point of the entangling surface.
These surfaces all reach the cosmological horizon (located at $\beta=0$) for $r/L=\pi/2$ with the first derivative vanishing. The integration constant has been chosen in a way so that we reach $\beta=\pi/2$ for $r/L=r^\star/L$. As pointed out in \cite{Geng:2019bnn}, computing the area of these surfaces always leads to the whole dS entropy and is independent of the value of $r^\star$. However, since the second derivative is non-vanishing on the UV slice, the integral will lead to different values if we introduce a UV cutoff\footnote{We thank Eva Silverstein for pointing us to the very interesting topic of cutoff (A)dS.}. We will place this hard Dirichlet cutoff on which the entangling surfaces end at
\begin{equation}
    r_c/L=\eps/L.
\end{equation}
More precisely, the entangling surfaces will not all go to $\beta=0$ anymore but depending on the position of the turning point $r^\star$, scan through all possible values of $\beta$ with the value of $\beta$ on the cutoff surface given by $\beta_\eps=\arcsin(\tan(r^\star/L)/\tan(\eps/L))$. This is already the case for the AdS spacetime with no cutoff present. As we will see, by solving the integral for the entanglement entropy, the entanglement entropy gets smaller for smaller intervals (larger values of $\beta_\eps$). The Dirichlet wall ''eats up" the entangling surfaces for increasing values of $\varepsilon$ due to the requirement $r^\star/L>\varepsilon/L$ (Fig. \ref{pic:geo}).

In order to present the entanglement entropy in a compact way, we switch to yet another parametrization for the entangling surfaces $r(\beta)$ in which the entangling surfaces minimize the Lagrangian
\beq
{\cal L}_{II} =L^{D-3}\,  \cos^{D-3}(\beta) \sin^{D-3} \left ( \frac{r}{L} \right ) \sqrt{(r')^2 +L^2\, \sin^2 \left ( \frac{r}{L} \right )\label{para2}
},
\eeq
and are given by
\beq
r(\beta)=L\,\arccot\left(\sin(\beta)/\tan(r^\star/L)\right).\label{solpara2}
\eeq
The entanglement entropy follows by computing the area of the minimal surfaces; evaluating the Lagrangian \eqref{para2} on the analytical solution \eqref{solpara2} and integrating from the cutoff surface at $\beta=\beta_\eps$ to the turning point of the entangling surfaces $\beta=\pi/2$ gives us the area 
\begin{align}
\int_{\beta_\eps}^{\pi/2}d\beta \mathcal L(r(\beta))&=\frac{L^{D-2}\sqrt{\pi}\,\Gamma(D/2-1)}{2\Gamma(D/2-1/2)}-\frac{L^{D-2}\,\text{}_2F_1[1/2,2-D/2,3/2,\frac{\sin(\beta_\eps)^2}{\sin(r^\star)^2+\cos(r^\star)^2\sin(\beta_\eps)^2}]}{\sqrt{\cos(r^\star)^2+\sin(r^\star)^2/\sin(\beta_\eps)^2}}\nonumber\\&\sim
\text{EE}_\text{dS}-\Delta(\eps,r^\star),\label{genEEdS}
\end{align}
where $\text{}_2F_1$ is the hypergeometric function $\text{}_2F_1(a,b;c;z)$. EE$_\text{dS}$ denotes the the full dS entropy which we get for the special cases $r^\star=0$ (studied in \cite{Gorbenko:2018oov}) or $\epsilon=0$ (studied in \cite{Geng:2019bnn}). We see that the entanglement entropy gets smaller for $\epsilon>0$ and $r^\star>0$. Since $r^\star$ is a bulk variable which does not have any obvious field theory interpretation, we want to eliminate it from the result. This may be done by using the analytical solution \eqref{solpara2} once more by calculating the position of the turning point ($\beta=\pi/2$) $r^\star=L\,\arctan\left(\frac{R\,\sin(\beta_\eps)}{\sqrt{L^2-R^2}}\right)$, where we also introduced the radius $R=L\,\sin(\eps/L)$ on the slice which is determined by evaluating the warpfactor for the position of the cutoff surface. With this, we finally arrive at
\begin{equation}
\Delta(\eps,\beta_\eps)=2\,L^{D-3}\,\sqrt{L^2-R^2\,\cos(\beta_\eps)^2}\,\text{}_2F_1\left(1/2,2-D/2,3/2,1-\frac{R^2\,\cos(\beta_\eps)^2}{L^2}\right).
\end{equation}
\subsection{Dirichlet walls and Entanglement Entropy in AdS}
The entanglement entropies of the preceding section may be interpreted in terms of the DS/dS correspondence. In this section we will focus on its parent, the AdS/CFT correspondence and mimic the calculation of the preceding section for AdS. In contrast to dS, AdS may be sliced in AdS, flat, or dS slicing. AdS$_{d}$ sliced AdS$_D$ follows from dS$_d$ sliced dS$_D$ by Wick rotation of both, the $D$-dimensional curvature constant and the $d\!=\!D\!-\!1$-dimensional curvature constant on the slice\footnote{For AdS$_d$ sliced AdS$_D$ the Lagrangian is $\mathcal L_{II}=L^{D-3}\sinh^{D-3}(\beta)\,\cosh^{D-3}\left(\frac rL\right)\,\sqrt{(r'(\beta))^2+L^2\cosh^{2}\left(\frac rL\right)}$, with analytical solution $r(\beta)=L\,\arctanh\left(\cosh(\beta)\,\tanh(r^\star/L)\right)$.}, while dS$_d$ sliced AdS$_D$ follows by only Wick rotating the $D$-dimensional curvature constant; the latter will be used in this work. In this spirit, the entangling surfaces are the solution to the equations of motion following from the Lagrangian
\begin{equation}
\mathcal L_I=L^{D-3}\cos^{D-3}(\beta)\,\sinh^{D-3}\left(\frac rL\right)\,\sqrt{1+L^2\,\sinh^{2}\left(\frac rL\right)\,(\beta')^2}.    
\end{equation}
It is not hard to find the solution to the equations of motion, given by
\begin{equation}
   \beta(r)=\arcsin(\tanh (r^\star/L)/ \tanh (r/L)).
\end{equation}
Analogous to the dS case, we introduce a hard radial cutoff at $r/L=\eps/L$, with the corresponding radius of the sphere on the cutoff surface given by $R=L\,\sinh(\eps/L)$. Note that the turning point of the entangling surface in the bulk at $r^\star$ is related to the position $\beta_\eps$ where the entangling surface ends on the Dirichlet wall by $\beta_\eps=\arcsin(\tanh(r^\star/L)/\tanh(\eps/L))$.

We may calculate the entanglement entropy by evaluating the Lagrangian for the analytical solution and integrating along the entangling surface to yield the minimal area
\begin{equation}
A=2\,L^{D-3}\int_{r^\star}^{\eps}dr\,\frac{\sinh(r/L)}{\cosh(r^\star/L)}\,\left(-1+\frac{\cosh(r/L)^2}{\cosh(r^\star/L)^2}\right)^{D/2-2}. \label{OSLag}   
\end{equation}

To solve this integral, it was convenient to switch variables by introducing the auxiliary variable $y^2=-1+\cosh(r/L)^2/\cosh(r^\star/L)^2$, which transforms \eqref{OSLag} to
\begin{equation}
    A=2\,L^{D-2}\,\int_{y(r^\star)}^{y(\eps)}dy\,\frac{y^{D-3}}{\sqrt{1+y^2}}=\frac{(R \cos (\beta_\eps))^{D-2}}{D-2} \, _2F_1\left(\frac{1}{2},\frac{D-2}{2};\frac{D}{2};-\frac{R^2 \cos ^2(\beta_\eps)}{L^2}\right).\label{genEEAdS}
\end{equation}
\subsection{Entanglement entropies for general intervals on the sphere}
\begin{figure}
    \centering
  \includegraphics[width=7.1cm]{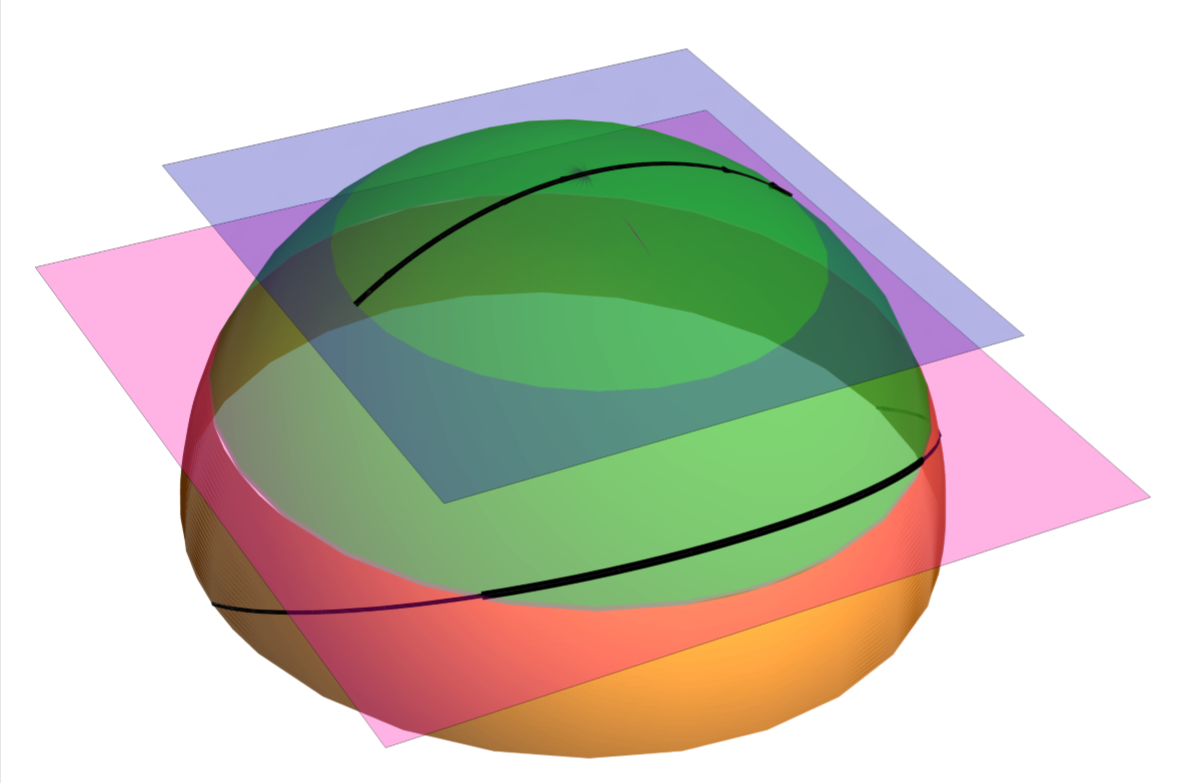}\hspace{0.5cm}  \includegraphics[width=4.5cm]{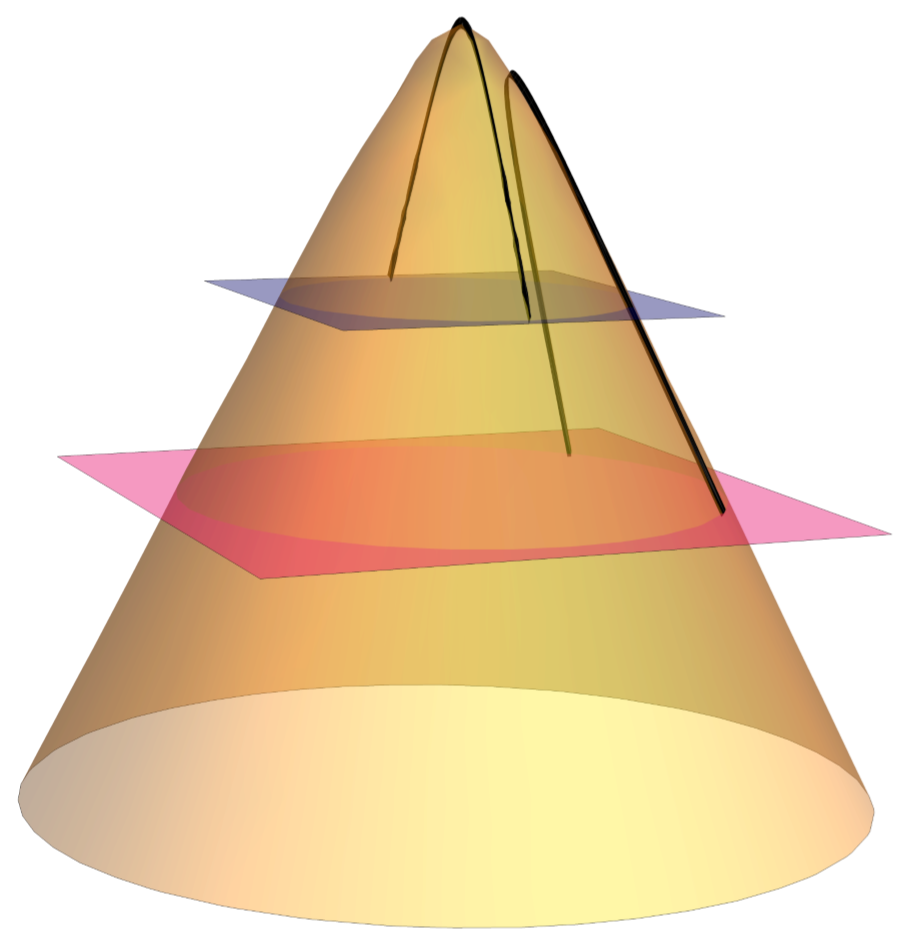}
       \caption{\textbf{Left:} The entangling surface for $r^\star/L=\pi/3$ -- $(\theta,r)$ are the polar and azimuthal angles, respectively, in the static patch of Euclidean dS$_3$ in presence of a cutoff $\epsilon$ (magenta surface). The cutoff surface restricts the entangling surface to the bolder line. We can rotate this surface by $\theta_0=\pi/3$ to bring it to the top of the sphere. If we draw a line through the ending points, we see that this corresponds exactly to a cutoff surface with radius $R_\text{eff}=R\,\cos(\beta_\epsilon(r^\star))$, which is depicted in blue. By rotating the surface on the circle, we see that the entangling surface exactly corresponds to the half-circle, i.e. the interval consists of antipodal points. The field theory lives on the circle on the magenta surface. \textbf{Right:} The analogous picture for Euclidean AdS$_3$. Note that the transformation consists of a spacetime rotation and a special conformal transformation.}\label{pic:geo}
\end{figure}

In equation \eqref{genEEdS} and \eqref{genEEAdS}, we derived expressions for the entanglement entropies for generic intervals in the presence of a Dirichlet wall which follow from the minimal area surfaces by
\begin{equation}
    S_\text{EE}=\frac{2\pi\,A}{\ell_P^{d-1}}.
\end{equation}
By varying the starting point of the entangling surfaces in the bulk $r^\star$, we are able to change the size of the interval on the sphere and thus calculate the entanglement entropy for subintervals. The case $r^\star=0$ corresponds to antipodal points on the sphere; smaller intervals on the sphere occur for larger values of $r^\star$. The radius $R$ of the sphere appears in equations \eqref{genEEdS} and \eqref{genEEAdS}, but only in combination with the cosine of the ending point of the entangling surfaces on the cutoff surface $R\,\cos(\beta_\eps)$; it is therefore useful to introduce an effective radius $R_\text{eff}(\beta_\eps)=R\cos(\beta_\eps)$. Introducing the effective radius makes it apparent that the entanglement entropies of the one parameter family still have the same form as the entanglement entropy of the special case $r^\star/L=0 \, (\beta_\eps=0)$, which is for AdS$_D$ and for $dS_3$ known in the literature \cite{Banerjee:2019ewu,Gorbenko:2018oov}; the entanglement entropies are decreasing for increasing $\beta_\eps$. For the sake of convenience, we list the results for the entanglement entropies in $D\!=\!3$ to $D\!=\!7$ and we label them with the dimension $d\!=\!D\!-\!1$ of the dual field theory. In the spirit of \cite{Gorbenko:2018oov}, we introduce $\eta$, with $\eta=1$ corresponding to AdS and $\eta=-1$ to dS. Furthermore, the (h) in expressions $\arcsinhh$ corresponds to the AdS case. The entanglement entropies read 
 \begin{align}
d=2:\ & S_\text{EE}(\beta_\eps)=\frac{4\,L\,\pi}{\ell_p}\,\arcsinhh\left(\frac{R_\text{eff}}{L}\right)\label{holoEEd2}\\
d=3:\ & S_\text{EE}(\beta_\eps)=\frac{4\,L\,\pi^2}{\ell_p^2}\,\eta\left(-L+\sqrt{L^2+\eta\,R_\text{eff}^2}\right)\label{holoEEd3}\\
d=4:\ & S_\text{EE}(\beta_\eps)=\frac{4\,\pi^2\,L}{\ell_p^3}\,\eta\left(R_\text{eff}\sqrt{\eta\,R_\text{eff}^2+L^2}-L^2\,\arcsinhh\left(\frac{R_\text{eff}}{L}\right)\right)\label{holoEEd4}\\
d=5:\ & S_\text{EE}(\beta_\eps)=\frac{4\,\pi^3\,L}{3\,\ell_p^4}\left(2L^3+(\eta\,R_\text{eff}^2-2\,L^2)\,\sqrt{L^2+\eta\,R_\text{eff}^2}\right)\label{holoEEd5}\\
d=6:\ & S_\text{EE}(\beta_\eps)=\frac{2\,\pi^3\,L}{3\,\ell_p^5}\!\left(\!R_\text{eff}\,\sqrt{L^2+\eta\,R_\text{eff}^2}\,(2\,\eta\,R_\text{eff}^2-\!3L^2)\!+\!3\,L^4\arcsinhh\!\left(\frac{R_\text{eff}}{L}\right)\!\right)\label{holoEEd6}\!.
\end{align}
\begin{figure}
    \centering
  \includegraphics[width=8cm]{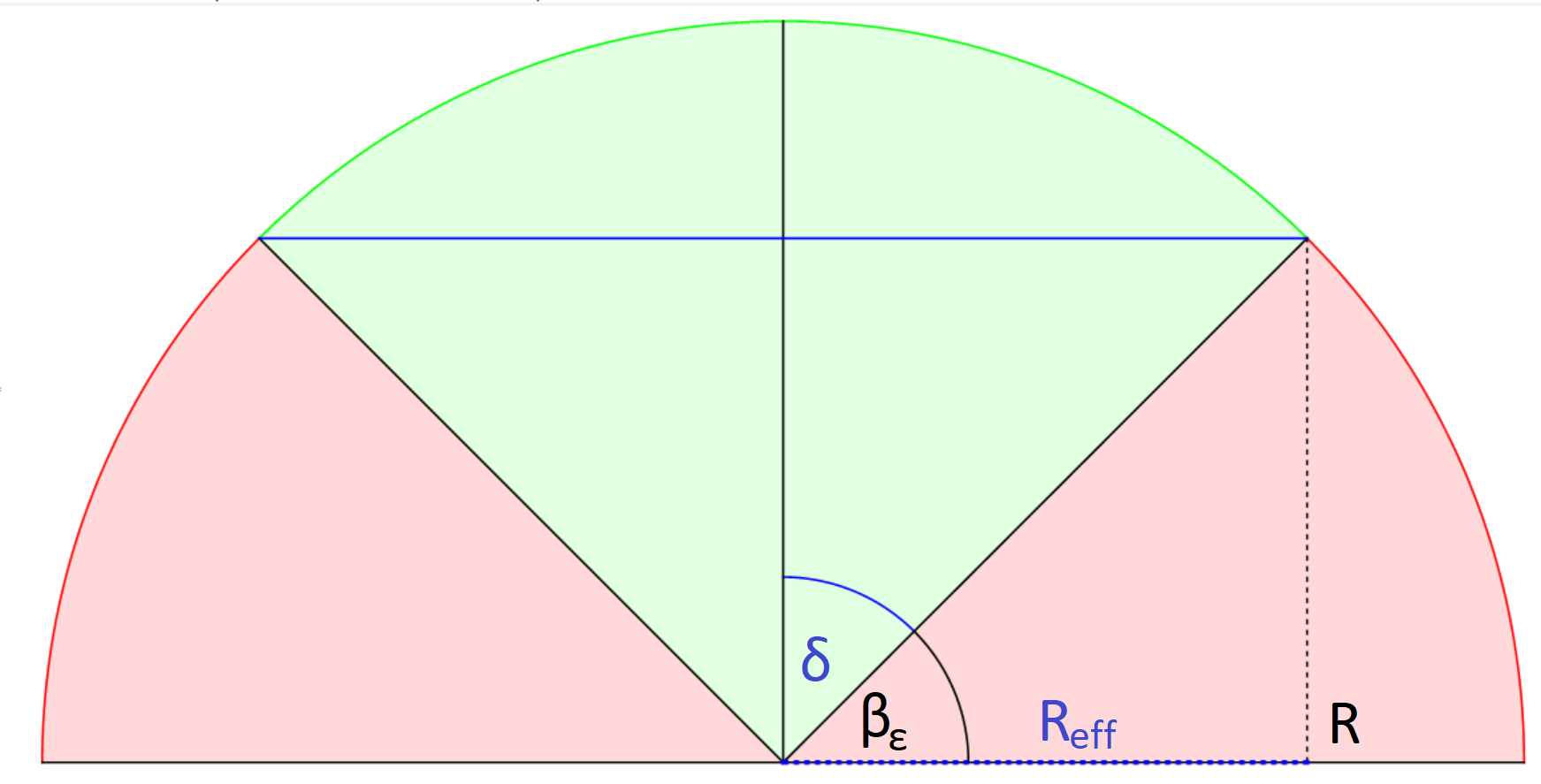}
       \caption{The interval under consideration on the circle of radius $R$ is depicted in green. The effective radius $R_\text{eff}=R\,\cos(\beta_\epsilon)$ corresponds by the definition of the cosine (dashed blue line) to the radius, where the points of the interval are antipodal.}\label{pic:geo2}
\end{figure}
The results are more straightforward if seen from a geometric perspective~(see figure \ref{pic:geo2} and \ref{pic:geo}). From the definition of the cosine, we see that the effective radius corresponds to the sphere where the endpoints of the interval are north and south pole. Without the cutoff, the one parameter family of entangling surfaces in the dS case (found in \cite{Geng:2019bnn}) are all just great circles on the sphere with the limiting surfaces $r^\star=0$ and $r^\star=\pi L/2$ corresponding to the equator and crossing over the north pole. Since they are all half-circles on the sphere, they all have the same area. If we introduce a cutoff surface at $r/L=\eps/L$, the surfaces all yield to a different area and thus to a different entanglement entropy. The Dirichlet wall cuts the one parameter family into surfaces of different length, depending on $r^\star$. As a result, we are able to calculate the entanglement entropy for different intervals on the circle. As shown in the graphic, those surfaces may be rotated along the sphere until they correspond to a half-circle again; the half-circle has the radius $R_\text{eff}=R\cos(\beta_\eps(r^\star))$. On this half-circle, the entangling surface corresponds to the entanglement entropy of two antipodal points; moving the cutoff surface up to the effective radius may also be done by following the $T\bar T$ trajectory. In the AdS case, this may be done by rotating the entangling surface up to the apex of the cone with a spacetime rotation and then applying a special conformal transformation to bring the entangling surface on the surface of the cone; these transformations map the points of a generic interval on a sphere with radius $R$ to antipodal points on a sphere with radius $R_\text{eff}$.

It is important to note that the angle $\beta_\eps$ measures how much the interval gets smaller compared to an interval of antipodal points on the sphere. The case $\beta_\eps=0$ corresponds to antipodal points. In order to measure the length of the interval, it makes sense to introduce the angle $\delta=\pi/2-\beta$, with $R_\text{eff}=R\cos\theta_\eps=R\sin\delta$. In order to further confirm our results, we expand the results for pushing the cutoff surface to the boundary. We reach the boundary for $R\to\infty$ (AdS) and $R=L$ (dS), respectively. Introducing the cutoff $\Lambda$, the entanglement entropies for AdS read
\begin{align}
 S^{d=2}_\text{EE}(\delta)&\!=\!\frac{4\,L\,\pi}{\ell_p}\,\left(\log\!\left(\frac{2\,\Lambda \sin(\delta)}{L}\right)+\frac{L^2}{4 \Lambda^2\,\sin(\delta)^2}+\mathcal O\left(\frac{1}{\Lambda^3}\right)\right)\label{holoEEd2a}\\
 S^{d=3}_\text{EE}(\delta)&\!=\!\frac{4\,L\,\pi^2}{\ell_p^2}\!\left(\Lambda\,\sin(\delta)-L+\frac{L^2}{2\Lambda\,\sin(\delta)}+\mathcal O\left(\frac{1}{\Lambda^3}\right)\right)\label{holoEEd3a}\\
 S^{d=4}_\text{EE}(\delta)&\!=\!\frac{4\,\pi^2\,L}{\ell_p^3}\!\left(\Lambda^2 \sin(\delta)^2-\frac 12 L^2-L^2\log\!\left(\frac{2\,\Lambda \sin(\delta)}{L}\right)+\frac{L^2}{4 \Lambda^2\,\sin(\delta)^2}\!+\!\mathcal O\!\left(\frac{1}{\Lambda^2}\!\right)\!\right)\label{holoEEd4a}\\
 S^{d=5}_\text{EE}(\delta)&\!=\!\frac{4\,\pi^3\,L}{3\,\ell_p^4}\!\left(\Lambda^3\,\sin(\delta)^3-\frac 32 \Lambda L^3\,\sin(\delta)+2L^3-\frac{9 L^4}{8\,\Lambda\,\sin(\delta)}+\mathcal O\left(\frac{1}{\Lambda^3}\right)\right)\label{holoEEd5a}\\
  S^{d=6}_\text{EE}(\delta)&\!=\!\frac{2\,\pi^3\,L^3}{3\,\ell_p^5}\!\left(\!\frac{2\Lambda^4\sin(\delta)^4}{L^2}-2\Lambda^2\sin(\delta)^2-\!\frac{7\,L^2}{4}+3L^2\log\!\left(\!\frac{2\Lambda \sin(\delta)}{L}\!\right)\!+\!\mathcal O\!\left(\frac{1}{\Lambda^2}\!\right)\!\right)\!.\label{holoEEd6a}
\end{align}
which matches the result of Casini, Huerta and Myers \cite{Casini:2011kv}. The results for $d=2$ match the well known field theory result for a subsystem $\ell$ in a system of length $L$ \cite{Holzhey:1994we,Calabrese:2004eu,Calabrese:2005zw}
\begin{equation}
S=\frac{c}{3}\,\log\left(\frac{L}{\pi\, a}\,\sin\left(\frac{\pi \ell}{L}\right)\right),
\end{equation}
with the cutoff $a$ ($a\to 0$).
In the dS case, $\Delta$ in \eqref{genEEdS} vanishes and we get back the full dS entropy as was observed in \cite{Geng:2019bnn}.

\subsection{Renormalization and generalized entanglement entropies}\label{genwald}
In general, if we consider entanglement entropies, we expect the result to be divergent, i.e. the leading divergence is the so-called area term
\cite{Nishioka:2009un,Casini:2003ix,Plenio:2004he,Cramer:2005mx,Das:2005ah}. In CFT calculations, however, we usually work with renormalized quantities instead of the bare ones since those quantities are universally well defined and still make sense when we take the continuum limit. The $T\bar T$ deformation, on the other hand, acts as UV regulator and all quantities are automatically finite. In principle, we could add an arbitrary amount of counterterms to the dual effective
field theory action but we will restrict ourselves to only considering the standard holographic counterterms~(\cite{deHaro:2000vlm,Skenderis:2002wp}). If we add counterterms to the field theory action, living on the cutoff slice, these finite counterterms will affect the result for the entanglement entropy (see for example \cite{Murdia:2019fax} for a discussion about this) \cite{Taylor:2016aoi,Cooperman:2013iqr,Emparan:1999pm,Faulkner:2013ana,Faulkner:2013ica}.
In the discussion of the next section, we will consider the renormalized stress tensor on the cutoff slice which can be derived by supplementing the gravitational action with counterterms in order to render it finite as explained in \cite{deHaro:2000vlm,Skenderis:2002wp}. Specifically, we are considering the action
\begin{equation}
S_\text{tot}=S_\text{EH}+S_\text{surf}+S_\text{ct},\label{actionS}
\end{equation}
with \begin{align}
    & S_\text{EH}=-\frac{1}{2\,\ell_p^{d-1}}\int\dd^{d+1}x\,\sqrt{g}\,\left(R^{(d+1)}-2\Lambda\right),\\
   & S_\text{surf}=-\frac{1}{\ell_p^{d-1}}\int\dd^dx\,\sqrt{\gamma}\,K,\\
    &S_\text{ct}=\frac{1}{2\,\ell_p^{d-1}}\int\dd^dx\,\sqrt{\gamma}\left(\!2\,c_1^{(d)}\frac{d-1}{L}+\frac{c^{(d)}_2\,L}{d-2}\,\tilde R+\frac{c_3^{(d)}\,L^3}{(d-4)\,(d-2)^2}\left(\!\tilde R_{ij}\tilde R^{ij}-\frac{d}{4\,(d-1)}\tilde R^2\right)\!\right)\!,\label{counter}
\end{align}
where $\tilde R,\,\tilde R_{ab}$ are the Ricci scalar- and tensor, respectively, on the boundary slice with the induced metric $\gamma$. Furthermore, the $c_{i}^{(d)}=1$ in case of: $c_1^{(d)}: d\ge 2$, $c_2^{(d)}: d\ge 3$ and $c_3^{(d)}: d\ge 5$ and $K$ is the extrinsic curvature.
In the derivation of the holographic entanglement entropy, we equate the partition function of both theories
\begin{equation}
    Z_\text{CFT,bare}[\gamma_{ij}]=\left.e^{I_\text{grav}[g_{ij}]}\right|_{r=r_c}.
\end{equation}
However, this is a statement about the \textit{bare} partition functions of both theories. If we renormalize the field theory partition function, we have to take into account the counterterms in the gravitational theory which act as higher curvature terms on the cutoff slice. Thus, we may take into account the counterterms on the cutoff slice by adding the contributions of the Wald entropy associated with the counterterms to the holographic entanglement entropy. The Wald entropy \cite{Wald:1993nt} is given by \cite{Jacobson:1993vj,Jacobson:1994qe,Brustein:2007jj}
\begin{equation}
    S_\text{Wald}=-2\pi\oint\dd^dx\,\frac{\delta\mathcal L}{\delta\tilde R_{abcd}}\,\hat\epsilon_{ab}\,\hat\epsilon_{cd},\label{wald}
\end{equation}
where $\hat\epsilon_{ab}$ are the binormals to the horizon. For pedagogical reasons, we rewrite the metric eq. \eqref{metricAdsdS} with $R(r)=L\,\sinhh(r/L), R\equiv R(r_c)$ and $\rho=\cos\phi $
\begin{equation}
    \dd s^2_\text{DS}=\dd r^2+ R^2(r)\,\left(-(1-\rho^2)\,\dd \tau^2+\frac{\dd \rho^2}{1-\rho^2}+\rho^2\,\dd\Omega_{d-2}\right).
\end{equation}
In these coordinates, we see that on the cutoff slice $r/L=r_c/L$ in the static patch, the $\hat\epsilon_{\tau\rho}$, are the binormals and we have to vary the Lagrangian in eq. \eqref{wald} with respect to $\tilde R_{\tau\rho\tau\rho}$ in order to find the Wald entropy.
For the counterterms given in eq. \eqref{counter}, we thus find on the slice $r/L=r_c/L$ for an entangling surface with $\rho_\epsilon=\cos(\beta_\epsilon)$ and $R_\text{eff}=R(r_c)\,\rho_\epsilon$
\begin{align}
    S_\text{W,ct}&=-\frac{2\pi}{\ell_p^{d-1}}R_\text{eff}^{d-2}\,\oint \sqrt{h}\,\left(\frac{c_2^{(d)}\,L}{d-2}+\frac{c_3^{(d)}L^3}{(d-4)(d-2)^2}\left(\tilde R-h^{ab}\tilde R_{ab}-\frac{2d}{4\,(d-1)}\tilde R\right)\right)\nonumber\\
    &=-\frac{4\, \pi^{(d+1)/2} \,R_\text{eff}^{d-2}}{(d-2) \,\ell_p^{d-1}\,\Gamma((d-1)/2)} \left(c_2^{(d)}\,  L-\frac{c_3^{(d)}\, L^3}{2\,(d-4)\, R_\text{eff}^2}\left(d-2\right)\right),\label{ctwald}
\end{align}
where $h_{ab}$ is the induced metric on the unit sphere and where we used that on the cutoff slice $\tilde R=d\,(d-1)/R_\text{eff}^2$ and $R_{ab}=(d-1)/R_\text{eff}^2\,h_{ab}$. 
Evaluating the expression in eq. \eqref{ctwald} for $3\le d\le 6$, we find
\begin{align}
    S^{d=3}_\text{W,ct}&=-\frac{4\,\pi^2\,L\,R_\text{eff}}{\ell_p^2}\label{wald1} \\
    S^{d=4}_\text{W,ct}&=-\frac{4\,\pi^2\,L\,R_\text{eff}^2}{\ell_p^3}\label{wald2} \\
    S^{d=5}_\text{W,ct}&=\frac{\pi^3\,L}{\ell_p^3}\left(-\frac{4\,R_\text{eff}^3}{5}+2\,R_\text{eff}\,L^2\right) \label{wald3} \\
    S^{d=6}_\text{W,ct}&=\frac{\pi^3\,L}{\ell_p^4}\left(-\frac{4\,R_\text{eff}^4}{3 }+\frac{4\,R_\text{eff}^2\,L^2}{3}\right)\label{wald4}.
\end{align}
Note that in $d=2$, we do not see a contribution from the counterterms to the entanglement entropy since the counterterm acts as a boundary cosmological constant.

\section{d-dimensional $T\bar T$ deformations in field theory}\label{ttfield}
In the second part of this work we take a closer look at the field theory side and compute the entanglement entropy for antipodal points in general dimensions in context of DS/dS. In order to find the entanglement entropies, we have to derive the analog of the higher dimensional $T\bar T$ like deformation for dS. As in the preceding section, we establish a notation in which the AdS and the dS case go hand in hand. We keep the derivations in this chapter short and refer the interested reader to \cite{Donnelly:2018bef,Caputa:2019pam,Hartman:2018tkw,Banerjee:2019ewu,Gorbenko:2018oov}.

\subsection{The d-dimensional deforming operator for holographic stress tensors}
We may extract the Brown-York stress tensor from the renormalized action eq. \eqref{actionS} by considering
\begin{equation}
   \delta S_\text{tot}=\frac 12\int\dd^dx\,\sqrt{\gamma}\ T^{ij}\,\delta \gamma_{ij},\label{EMtensor2}
\end{equation}
where $\gamma$ is the induced metric on the cutoff slice.
The stress tensor of the boundary field theory $T_{ij}^\text{bdy}$ is related to the bulk stress tensor by rescaling $T^\text{BY}_{ij}=r_c^{d-2}\,T^\text{bdy}_{ij}$. This is also true for the metric of the CFT: $g_{ij}(r=r_c,x)=\gamma_{ij}(x)=r_c^2\,\gamma_{ij}^\text{bdy}(x)$. For a complete dictionary on the cutoff slice see \cite{Hartman:2018tkw}. In the following discussion, we will set $r_c=1$.
The holographic stress tensor dual to a $d$-dimensional field theory may be expressed in terms of the extrinsic curvature $K_{ij}$ and the induced quantities on the boundary slice: the metric $\gamma_{ij}$, the Einstein tensor $\tilde G_{ij}$, the Riemann tensor $\tilde R_{ijkl}$, the Ricci tensor $\tilde R_{ij}$ and the Ricci scalar $\tilde R$~\cite{deHaro:2000vlm,Balasubramanian:1999re,Hartman:2018tkw,Banerjee:2019ewu}. The renormalized stress tensor consists of two components $T^\text{ren}_{ij}[\gamma]=T_{ij}[\gamma]+C_{ij}[\gamma]$, the standard holographic stress tensor on the cutoff surface $r=r_c$, $T_{ij}$, and the corresponding curvature contributions of the counterterms eq. \eqref{counter}, denoted by $C_{ij}$. In order to remove the divergences in the action eq. \eqref{actionS}, we add counterterms eq. \eqref{counter} -- which are scalar quantities -- to the action. The divergences arise when we push the cutoff surface to the AdS boundary. On the cutoff surface, the stress tensor is automatically regularized and reads
\begin{align}
    T_{ij}=&\frac{1}{8 \pi G_N}\,\left(K_{ij}-K\,\gamma_{ij}-c_d^{(1)}\,\frac{d-1}{L}\,\gamma_{ij}+\frac{c_d^{(2)}L}{d-2}\,\tilde G_{ij}\right.\nonumber\\
    &\left.\ +\frac{c_d^{(3)}L^3}{(d-4)(d-2)^2}\,\left(2\,\left(\tilde R_{ijkl}-\frac 14\,\gamma_{ij}\,\tilde R_{kl}\right)\,\tilde R^{kl}-\frac{d}{2\,(d-1)}\,\left(\tilde R_{ij}-\frac 14\,\tilde R\,\gamma_{ij}\right )\tilde R\right.\right.\nonumber\\
    &\ \left. \left. -\frac{1}{2\,(d-1)}\left(\gamma_{ij}\,\Box \tilde R+(d-2)\,\nabla_i\,\nabla_j\tilde R\right)+\Box\tilde R_{ij}\right)\right),\label{EMtensor}
\end{align}
with $\ell_P^{d-1}=8\pi\,G_N$.  With eq. \eqref{counter} and eq. \eqref{EMtensor2} the curvature dependent counterterms give thus rise to the contribution (in $d\ge 3$)~\cite{deHaro:2000vlm,Skenderis:2002wp}
\begin{align}
C_{ij}=&-\frac{1}{8 \pi G_N}\,\left(c_d^{(2)}\,\tilde G_{ij}+c_d^{(3)}\,b_d\,\left[2\left(\tilde R_{ikjl}-\frac14\,\gamma_{ij}\,\tilde R_{kl}\right)\,\tilde R^{kl}-\frac{d}{2(d-1)}\,\left(\tilde R_{ij}-\frac14\,\tilde R\,\gamma_{ij}\right)\tilde R\right.\right.\nonumber\\
&\left.\left. -\frac{1}{2\,(d-1)}\left(\gamma_{ij}\,\Box \tilde R+(d-2)\,\nabla_i\nabla_j\tilde R\right)+\Box\tilde R_{ij}\right]\right),\label{EMcounter}
\end{align}
with $b_d=l^2/((d-4)(d-2))$. The $c^{(d)}_i$'s are non-zero in case of: $c^{(d)}_2=1$ for $d\ge 3$ and $c^{(d)}_3=1$ for $d\ge 5$.
Deforming a theory by a local operator $X$ results, on the level of the classical action, in
\begin{equation}
    \frac{\partial S}{\partial \lambda}=\int d^dx\,\sqrt{\gamma} X,
\end{equation}
with $\lambda$ being the size of the deformation. In $d=2$, the $T\bar T$ deformation satisfies the factorization property~\cite{Zamolodchikov:2004ce}
\begin{equation}
    \langle T\bar T\rangle=\frac 18\,\left(\langle T^{ij}\rangle\langle T_{ij}\rangle-\langle T_i^i\rangle^2\right).\label{facto}
\end{equation}
In general, this would no longer be true in higher dimensions. However, since we are working in a CFT at large $N$, the factorization property is still valid at $d>2$ \cite{McGough:2016lol,Hartman:2018tkw}.

In order to derive the deforming operator $X_d=-\frac{1}{d\,\lambda_d}T^i_i$, we derive the trace flow equation for the holographic stress tensor using Einstein's equations. This is accomplished by using the general form of the holographic stress tensor \eqref{EMtensor} and then by using the Hamilton constraint in appendix \ref{ddimdefo}. The d-dimensional expression is given by
\begin{align}
X_d=\!\left(T_{ij}\!+\!\frac{\alpha_d}{\lambda_d^{\frac{d-2}{d}}}C_{ij}\!\right)^{\!2}\!\!\!-\!\frac{1}{d-1}\left(T^i_i\!+\!\frac{\alpha_d}{\lambda_d^{\frac{d-2}{d}}}C_i^i\!\right)^{\!2}\!\!+\!\frac1d \frac{\alpha_d}{\lambda_d^{\frac{2(d-1)}{d}}}\!\left(\frac{d-2}{2}\mathcal R^{(d)}\!+\!C^i_i\!\right)\!+\!\frac{(d-1)(\eta-1)}{4\,d\,\lambda_d^2},\label{defo}
\end{align}
where $\eta=1$ corresponds to the AdS case and $\eta=-1$ to the dS case, respectively. 
Furthermore, the $\alpha_d$ are dimensionless numbers and correspond to the number of degrees of freedom in the field theory. We denote the coupling of the deformations by $\lambda_d$.
The parameters of the field theory are related to the parameters on the gravity side by~\cite{Banerjee:2019ewu}
\begin{equation}
    \lambda_d=\frac{\ell_P^{d-1}\,L}{2d},\quad \alpha_d=\frac{L^{2(d-1)/d}}{(2d)^{\frac{d-2}{d}}\,(d-2)\,\ell_P^{\frac{2\,(d-1)}{d}}},\quad L^2=2d\,(d-2)\,\alpha_d\,\lambda^{2/d}_d.\label{gravityfieldtheory}
\end{equation}
In a two-dimensional CFT, the central charge $c$ is related to the bulk quantities as \cite{Brown:1986nw}
\begin{equation}
    c=\frac{12 \pi L}{\ell_P}.
\end{equation}
For example in $d=2$ dimensions the deforming operator reads \begin{equation}
    X_2=T_{ij}^2-\frac{1}{d-1}(T^i_i)^2+\frac{1}{2\lambda}\,\frac{c}{24\pi}\,R+\frac{\eta-1}{8\lambda^2},
\end{equation}
which matches the result of \cite{Gorbenko:2018oov}.

\subsection{Sphere partition functions and entanglement entropy}
Let us consider a generic seed CFT in $d$-dimensions at large central charge on a sphere with radius $R$. Our goal is to compute the exact sphere partition function $Z_{S^d}$. From the sphere partition function, it is straightforward to calculate the entanglement entropy for antipodal points on the sphere as was outlined by \cite{Donnelly:2018bef}. What we are interested in is the change of the partition function in response to deformations of the sphere. As argued in \cite{Donnelly:2018bef,Banerjee:2019ewu} and since changes of the metric manifest in the vacuum expectation value of the stress tensor, the symmetries on the sphere dictate 
\begin{equation}\langle T_{ij}\rangle=\omega_d(R)\,\gamma_{ij}\label{symmetrysphere}
\end{equation} 
we can write the deformation of the $d-$dimensional sphere partition function as
\begin{equation}
R\frac{\partial}{\partial R}\,\log Z_{S^d}=-d\,\int d^dx\,\sqrt{\gamma}\,\omega_d(R),\label{partf}
\end{equation}
from which we can compute the entanglement entropy using the replica trick. It now becomes apparent why we chose dS slicing for (A)dS in the first place: the dS ground state corresponds to the Euclidean path integral on the sphere $S^d$.

We may apply the replica trick by considering the $n$-folded cover of the sphere of radius $R$ \cite{Donnelly:2018bef,Banerjee:2019ewu}
\begin{equation}
    ds^2=R^2\left(d\theta_1^2+\sum_{i=2}^{d-1}\prod_{j=1}^{i-1}\cos(\theta_j)^2\,d\theta_i^2+n^2\,\prod_{j=1}^{d-1}\cos(\theta_j)^2\,d\theta_d^2 \right),\label{nfoldcover}
\end{equation}
with $\theta_j\in[-\pi/2,\pi/2]$ for $j=1,\ldots,d-1$ and $\theta_d\in[0,2\pi]$. For simplicity, we set $d=2$ for now which reduces \eqref{nfoldcover} to
\begin{equation}
    ds^2=R^2\,(d\theta^2+n^2\cos(\theta)^2 d\phi^2),
\end{equation}
with $\phi\in[0,2\pi]$ and $\theta\in[-\pi/2,\pi/2]$. The angle $\theta$ is chosen so that it corresponds to the angle $\theta$ in the gravitational theory; it is the azimuthal angle on the sphere and the antipodal points $\theta=-\pi/2$ and $\theta=\pi/2$ correspond to the north and south pole of the sphere. Since the entangling surface consists of two antipodal points we may -- due to the rotational symmetry -- continuously vary $n$ which allows us to compute the entanglement entropy with \begin{equation}
S_\text{d,EE}=\left(1-\frac{R}{d}\frac{d}{dR}\right)\log Z_{S^d}.\label{EEreplica}
\end{equation}
In order to calculate the sphere partition function \eqref{partf}, we compute the expression for $\omega_d$ using the flow equation $\langle T^i_i\rangle=-d \,\lambda_d\,\langle X_d\rangle$. With help of the deforming operator $X_d$ (defined in eq. \eqref{defo}), we end up with a quadratic equation for $\omega_d$. To derive an explicit expression for $\omega_d$, we have to evaluate the stress energy tensor and the counterterms for a $d$-dimensional sphere of radius $R$. The quadratic equation yields a positive and a negative solution so there are two possible signs for the $T\bar T$ deformation.
For now, we will denote the signs of the square root simply by $\bm{s}$.
In the case $d=2$ all the $c's$ are zero and it is straightforward to check that
\begin{equation}
 \omega_2=\frac{1+\bm{s}\sqrt{\eta+\frac{c\,\lambda_2}{3\pi\,R^2}}}{4\,\lambda_2}.\label{omega2}   
\end{equation}
Since we are working in a large $N$ CFT where the coupling of the deformation $\lambda^{2/d}$ is small but $N \lambda^{2/d}$ is finite, we may write (in $d>2$) the expansion parameter as $t_d=\alpha_d \lambda^{2/d}$. With this, we find the expression for the sphere partition function in $d>2$ to be
\begin{align}
\omega_{d>2}=&\frac{(-1+d)}{4\,d\,R^4\,\lambda_d}\left(\!2\,(d-2)\,d\,R^2\,\lambda^{2/d}\,c_d^{(2)}\,\alpha_d-(d-2)^2\,d^2\,\lambda^{4/d}\,c_d^{(3)}\,\alpha_d^2\right.\nonumber\\ 
&\left.+ 2R^3\,\left(R+\bm{s}\,\sqrt{\eta\,R^2+2\,(d-2)\,d\,\lambda^{2/d}\alpha_d}\right)\right)\label{omegaeq}
\end{align}
From the expression for $\omega_d$, it is straightforward to calculate the entanglement entropy. The procedure goes as follows: we are taking $\omega$ from eq. \eqref{omega2} and \eqref{omegaeq}, respectively and inserting them into eq. \eqref{partf}; this yields an expression for the $R$-derivative of the partition function. To obtain the entanglement entropy, we must first integrate the expression and plug the result into eq. \eqref{EEreplica}. However, this integration results in an integration constant that must be fixed before proceeding. We may fix the integration constant either in the IR or the UV region of the theory.
In $d=2$, we may follow \cite{Gorbenko:2018oov} and fix the integration constant by matching the partition function for AdS in the $R^2/\lambda_2\to\infty$ limit to the CFT partition function
\begin{equation}
    \log Z_\text{CFT}(R)=\frac c3\,\log\frac Ra.
\end{equation}
The integration constant for the dS case is obtained by matching the partition function in the $R^2/\lambda\to 0$ case to the AdS partition function\footnote{Note that $a$ is an arbitrary cutoff scale and not determined by the CFT itself. The value of the UV scale may be adjusted later. This implies that we have not fixed the integration constant before fixing the UV scale.  However, if we consider cutoff independent quantities as $\mathcal S_\text{R,EE}$ this constant does not contribute.}.
However, the CFT partition function is not known in $d>2$ and we chose to follow  \cite{Donnelly:2018bef}, which fixes the integration constant by demanding the logarithm of the partition function to vanish for $R^2/\lambda\to0$. This leads to a trivial theory in the UV. Note that this is only possible in presence of the UV cutoff since the theory does not change as a function of the scale at arbitrarily short distances anymore.

There is a different way to extract the universal quantities of the entanglement entropy as the approach we took in section \ref{genwald} based on \cite{Taylor:2016aoi,Cooperman:2013iqr}.
In analogy to \cite{Gorbenko:2018oov,Banerjee:2019ewu,Liu:2012eea,Liu:2013una}, we also compute the cutoff independent renormalized entanglement entropy\footnote{In contrast to \cite{Banerjee:2019ewu}, we follow the convention of \cite{Liu:2012eea,Liu:2013una} with the double factorial.}
\begin{equation}
    \mathcal S_\text{R,EE}(R)=\begin{cases}
    \frac{R}{(d-2)!!}\,R\,\frac{d}{dR}(R\,\frac{d}{dR}-2)\ldots(R\,\frac{d}{dR}-(d-2))\,S_\text{EE}\quad\quad\quad\quad\text{d even},
    \\ \frac{R}{(d-2)!!}(R\,\frac{d}{dR}-1)(R\,\frac{d}{dR}-3)\ldots(R\,\frac{d}{dR}-(d-2))\,S_\text{EE}\quad\quad\text{d odd.}\end{cases}\label{runningC}
\end{equation}

\section{Entanglement Entropy from field theory in general dimensions}
After reviewing the methods of how to compute the entanglement entropies for a $T\bar T$ deformed field theory, we are able to derive the expressions for the entanglement entropy in general dimensions. We eventually compare the expressions derived from field theory with the ones we computed on the gravity side in section \ref{EEholography}.
\subsection{$d=2$}
Let us reproduce the cases which are known in the literature so far. The results for the entanglement entropy for a $T\bar T$ deformation in a two-dimensional CFT are given in \cite{Donnelly:2018bef} (AdS) and \cite{Gorbenko:2018oov} (dS), respectively.

We have
\begin{equation}
        \omega_2=\frac{1+\bm{s}\sqrt{\eta+\frac{c\,\lambda_2}{3\pi\,R^2}}}{4\,\lambda_2}\label{omegad2}
\end{equation}
where we denote the sign of the square root of the $T\bar T$ deformation by $\bm{s}$. With $\omega_2$ at hand, we may compute the partition function of the deformed CFT using eq. \eqref{partf}. As argued in the previous section we choose the integration constant so that $\log Z_{S^2}(R=0)=0$. This yields
\begin{equation}
    \log Z_{S^2}=-\frac{1}{3\lambda_2}\left(c\,\bm{s}\,\arcsinhh\left(\frac{\sqrt{3\,\pi}\,R}{\sqrt{c\,\lambda_2}}\right)\,\lambda_2+\eta\,R\,\left(3\,R\,\pi+\bm{s}\,\sqrt{3\,\pi}\sqrt{\eta\,3\,R^2\,\pi+c\,\lambda_2}\right)\right).
\end{equation}
The entanglement entropy for two antipodal points on the sphere follows from the partition function via eq. \eqref{EEreplica} (for the negative sign of the square root)
\begin{equation}
    S_{EE}=
    \frac{c}{3}\,\arcsinhh\left(\frac{\sqrt{3\,\pi}\,R}{\sqrt{c\,\lambda_2}}\right).\label{EEd2}
\end{equation}
In two dimensions, we may calculate the cutoff independent renormalized entanglement entropy immediately from the knowledge of the derivative of the partition function.
 Plugging \eqref{omegad2} into eq. \eqref{partf} and combining eq. \eqref{EEreplica} and \eqref{runningC}, we find the renormalized entanglement entropy which plays the role of the running $\mathcal C$-function in RG flow as
\begin{equation}
    S_\text{R,EE}=
    c\left(9\,\eta+\frac{3\,c\,\lambda_2}{R^2\,\pi}\right)^{-1/2}.
\end{equation}
   \subsubsection*{Comparison to the result from holography}
The entanglement entropy from holography is given by eq. \eqref{holoEEd2}. In order to compare to the field theory results, we use the dictionary relating the holography parameters with the field theory ones. This is done by  $4\pi l/\ell_p=c/3$ and $c\,\lambda_2=3\,\pi\,L^2$ 
\begin{equation}
S_\text{EE}=\frac c3\,\arcsinhh\left(\frac{\sqrt{3\,\pi}\,R}{\sqrt{c\,\lambda_2}}\right),
\end{equation}
with the corresponding renormalized entanglement given by the $R$-derivative $\mathcal C=dS/dR$
\begin{equation}
   S_\text{R,EE}= c\left(9\,\eta+\frac{3\,c\,\lambda_2}{R^2\,\pi}\right)^{-1/2}.
\end{equation}
In $d=2$ dimensions, we find the entanglement entropy from field theory matches the entanglement entropy from holography exactly for $\eta=1,\bm{s}=-1$ (AdS) and $\eta=-1,\bm{s}=-1$ (dS).
\subsection{$3\le d\le6$}
The calculation for higher dimensions follows analogous to the calculation in
$d = 2$: compute
the  corresponding  partition  function  by  integrating  the  corresponding  \eqref{omegaeq},  use  the  partition function to compute the entanglement entropy according to \eqref{EEreplica} and eliminate scheme-dependent finite counterterms by differentiating using the prescription of  \eqref{runningC}. To avoid redundancy, we have moved the calculation to appendix \ref{EEhigherdim} and only display the results here. The bare holographic entanglement entropies in $3\le d\le 6$ match the field theoretic results only up to area terms, which are removed if we compute the renormalized entanglement entropies by taking derivatives with respect to $R$ according to the prescription of \cite{Liu:2012eea,Liu:2013una}. The reason for this discrepancy is due to the fact that we took counterterms into account in our field theory calculation which should also effect the gravity calculation since the counterterms are added on the cutoff slice. The area terms which appear with a negative sign in the field theory calculation -- and are thus subtracted from the result -- can be traced back to the counterterms. Since the $T\bar T$ deformation acts as a UV regulator, all quantities are already finite, especially the usually divergent leading area term contributions. However, if we take the contributions of the counterterms \eqref{counter} also in the gravity calculation into account, we see an exact match. In the following, we match the entanglement entropies obtained from the field theory calculation with the entanglement entropies obtained from holography (taking counterterms on the gravity side into account). Note that these expressions lack the so-called area terms which are removed by the renormalization
    \begin{align}
& S^{d=3,}_{\text{QFT}}=S^{d=3}_{\text{holo}}=\frac{4\,\pi^2\,t_3}{\lambda_3}\left(- R-\eta\,\sqrt{6t_3}+\eta\,\sqrt{\eta\,R^2+6\,t_3}\right) \left(\sqrt{6}-\frac{6 \sqrt{t_3}}{\sqrt{\eta\, R^2+6\, t_3}}\right)\\
 &S^{d=4}_{\text{QFT}}\!=S^{d=4}_{\text{holo}}\!=\frac{8\pi^2\,t_4}{\lambda_4}\left(R\left(-R+\eta\,\sqrt{\eta\, R^2+16\,t_4}\right)-16\,\eta\,t_4\,\arcsin(h)\left(\frac{R}{4\,\sqrt{t_4}}\right)\right)\\
 &S^{d=5}_{\text{QFT}}\!=S_{\text{holo}}^{d=5}\!=\!\frac{4\pi^3\,t_5}{\lambda_5}\! \left(\!-R^3\!+45 R\,t_5\!+\eta R^2\sqrt{\eta R^2+30 t_5}+\!60t_5\!\left(\!\sqrt{30\,t_5}\!-\!\sqrt{\eta R^2+30 t_5}\right)\!\right) \\
& S_{\text{QFT}}^{d=6}\!=S_{\text{holo}}^{d=6}\!= -\frac{16\,\pi^3\,t_6}{3\lambda_6}\left(R\left(R^3-48\,R\,t_6-\eta\,R^2\sqrt{\eta\,R^2+48\,t_6}+72\,t_6\,\sqrt{\eta\,R^2+48\,t_6}\right)\right.\nonumber\\
    &\ \quad\quad\quad\quad\quad\quad\left.-3456\,t_6^2\,\arcsinhh\left(\frac{R}{4\,\sqrt{3\,t_6}}\right)\right).\label{EEholoEF}
   \end{align}
 All the entanglement entropies calculated on the field theory side exactly match the entanglement entropies calculated on the gravity side of the duality for the negative sign of the square root. Furthermore, the entanglement entropies dual to the gravity theory in AdS match the entanglement entropies calculated in \cite{Banerjee:2019ewu}. 
\section{Conclusions}
In this work, we calculated the entanglement entropies for generic intervals in a (A)dS spacetime in presence of Dirichlet wall. This hard radial cutoff chops off the asymptotic UV region of the gravitational theory which is proposed to be the holographic dual to a CFT deformed by the irrelevant $T\bar T$ operator. Starting from one parameter families of analytical solutions for the entangling surfaces in (A)dS$_D$, we derived the associated entanglement entropies using the recipe of Ryu and Takayanagi. The entanglement entropies for antipodal points in (A)dS were already known in the literature \cite{Donnelly:2018bef,Gorbenko:2018oov,Banerjee:2019ewu}. Surprisingly, we may write the entanglement entropies for generic intervals formally in the same way as the already known results by introducing an effective radius $R_\text{eff}=R\,\cos(\beta_\eps)$. The basic definition of the cosine shows that the effective radius corresponds to a sphere where the endpoints of the interval are antipodal. Geometrically, this corresponds to the scenario where we follow the $T\bar T$ trajectory (move the cutoff inwards for intervals smaller than antipodal points) until the points of the generic interval are antipodal on a sphere with radius $R_\text{eff}$. In the dual field theory this means, we may compute the entanglement entropy of generic intervals on the sphere by the sphere partition function as explained in \cite{Donnelly:2018bef}, if we follow the $T\bar T$ trajectory. Note that in the AdS case this corresponds to a rotation in the spacetime followed by a special conformal transformation which brings the interval to antipodal points on the circle with radius $R\,\cos(\beta)$, as was illustrated in figure \ref{pic:geo}.
In the limit of pushing the Dirichlet cutoff to the boundary, we find agreement with the results of Casini, Huerta and Myers \cite{Casini:2011kv} (see eq. \eqref{holoEEd2a}-\eqref{holoEEd6a}). The authors of \cite{Chen:2018eqk} calculated perturbative corrections in the $T\bar T$ coupling to the entanglement entropy in a two dimensional field theory for finite intervals. They found that the leading order agrees with \cite{Casini:2011kv} while the first order corrections in the $T\bar T$ coupling vanish, as we see in \eqref{holoEEd2a}.
In the second part of this paper, we derived the entanglement entropies for antipodal points for a $d$-dimensional field theory in context of DS/dS holography. The $T\bar T$ deformations play an interesting role in DS/dS holography since they provide a mechanism to better understand the possible CFT dual to dS \cite{Gorbenko:2018oov}: with the help of $T\bar T$ deformations, we move the boundary inwards to the IR region; in the IR region the AdS and dS spacetimes are indistinguishable and the AdS/CFT-correspondence provides us with a CFT dual. In the IR, we may trigger the flow by using the $T\bar T$ deformation with the opposite sign. This time, we use the $T\bar T$ deformation, derived for the dS trajectory and we are able to move the boundary back to its original place. In this way, we are able to ``grow back" the spacetime we previously cut off but with a different sign for the cosmological constant. In section \ref{ttfield}, we extended the work of \cite{Gorbenko:2018oov} and derived the $T\bar T$ deformation in the context of DS/dS in general dimension. Compared to the deformation in AdS/CFT, the deformation gets an extra contribution proportional to the cosmological constant and the dimension of the field theory. We furthermore derived the (renormalized) entanglement entropies in general dimensions, which match our results derived from the gravitational theory perfectly. In particular, we worked out the contributions of the counterterms on the UV slice on the entanglement entropy in the gravitational theory by considering the Wald entropy associated with the counterterms. These contributions match exactly the contributions of the counterterms in our field theory calculations. The results thus shed light on the seeming mismatch of the entanglement entropies observed in \cite{Banerjee:2019ewu}. As the authors observed correctly, both \textit{bare} entanglement entropies match. However, if we switch on counterterms in the field theory calculation, the Ryu-Takayanagi prescription does not give the correct answer anymore; rather, we have to take the corrections of the counterterms into account by considering their Wald entropy.

\section*{Acknowledgements}
The author thanks  Se\'an Gray, Eva Silverstein and Andreas Karch for comments on the manuscript and Hao Geng and Andreas Karch for fruitful collaboration on related topics. Special thanks to Andreas Karch for numerous insightful discussions about entanglement entropy, cheering me up after impasses and eventually encouraging me to publish this work on my own.
The author gratefully acknowledges financial support by the  \textit{Fulbright Visiting Scholar Program}, which is sponsored by the US Department of State and the German-American Fulbright Commission in 2018 and  by the DAAD (German Academic Exchange Service) for a \textit{Jahresstipendium f\"ur Doktorandinnen und Doktoranden} (One-Year Research grant for doctoral candidates) in 2019. Last but not least, the author thanks Violeta and Renly for unconditional support.
   \appendix
\section{$d$-dimensional $T\bar T$ deformation in DS/dS}\label{ddimdefo}
The the radial Einstein equation for a $(d+1)-$dimensional gravitational theory in with metric \eqref{Adsds} in presence of a Dirichlet wall reads in terms of the extrinsic curvature $K_{ab}$
\begin{equation}
K^2-K^{ij}K_{ij}-\eta\,\frac{d(d-1)}{L^2}-\tilde R^{(d)}=0,\label{flow}
\end{equation}
with the induced $d$-dimensional Ricci scalar $\tilde R^{(d)}$. We may write the trace of the energy momentum tensor \eqref{EMtensor} together with the counterterms \eqref{EMcounter} as
\begin{equation}
    T_i^i+\frac{\alpha_d}{\lambda_d^{\frac{d-2}{d}}}\,C_i^i=\frac{d-1}{\ell_P^{d-1}}\left(K-\frac{d}{L}\right),
\end{equation}
which we may solve for the extrinsic curvature $K$ by considering the specific combination \begin{align}&\!\ell_p^{2(d-1)}\!\left(\!T_{ij}\!+\!\frac{\alpha_d}{\lambda_d^{\frac{d-2}{d}}}C_{ij}\!\right)^{\!2}\!\!\!-\frac{\ell_p^{2(d-1)}}{d-1}\!\left(\!T^i_i\!+\!\frac{\alpha_d}{\lambda_d^{\frac{d-2}{d}}}C_i^i\!\right)^{\!2}\!\!=\!\left(\!K_{ij}\!-\!K\,\gamma_{ij}-\frac{d-1}{L}\gamma_{ij}\!\right)^{\!2}\!-\!(d-1)\!\left(\!K\!-\!\frac{d}{L}\!\right)^{\!2}\nonumber\\
&=K^{ij}K_{ij}+(d-2)\,K^2+2\,K\,\frac{(d-1)^2}{L}+\frac{d\,(d-1)^2}{L^2}-(d-1)\left(K^2-2\, K\,\frac{d}{L}+\frac{d^2}{L^2}\right)\nonumber\\
&=K^{ij}K_{ij}-K^2+K\,\frac{2d-1}{L}-\frac{d\,(d-1)}{L^2}.
\end{align}
The deforming operator \eqref{defo} follows immediately by using eq. \eqref{flow}. Note that in $AdS$ the term $\sim d(d-1)/L$ cancel, while in dS they have the same sign and lead to an extra contribution to the deforming operator.
\section{Entanglement entropies from field theory}\label{EEhigherdim}
In this section, we present the computation to the results quoted in eq. \eqref{EEholoEF}. Since the computation is very repetitive, we focus on displaying the relevant steps.
\subsection{$d=3$}
  The procedure in $d=3$ is very similar to the case $d=2$; we may read off $\omega_3$ from eq. \eqref{omegaeq}
\begin{equation}
    \omega_3=\frac{R^2+3\,t_3+R\,\bm{s}\,\sqrt{\eta\,R^2+6\,t_3}}{3\,R^2\lambda_3}.
    \end{equation}
It is straightforward to determine the corresponding partition function, given by
    \begin{equation}
    \log Z_{S^3}=-\frac{2\pi^2\,(R^3+9\,R\,t_3+\eta\bm{s}\,(\eta\,R^2+6\,t_3)^{3/2})}{3\,\lambda_3}+\eta\,\bm{s}\frac{4\sqrt{6}\,\pi^2\,t^{3/2}}{3\,\lambda_3}.
\end{equation}
The second term is chosen to ensure $\log Z(R=0)=0$.
    Finally, we find with $\eta^2=1$ and eq. \eqref{EEreplica} and the negative sign of the square root
    \begin{equation}
        S_{EE}=\frac{4\,\pi^2\,t_3}{\lambda_3}\left(- R-\eta\,\sqrt{6t_3}+\eta\,\sqrt{\eta\,R^2+6\,t_3}\right).
    \end{equation}
The scheme independent renormalized entanglement entropy is obtained from the entanglement entropy by using \eqref{runningC} and reads in $d=3$ dimensions
   \begin{equation}
S_{\text{R,EE}}=\frac{4\, \pi^2\,\eta\,t_3^{3/2}}{ \lambda_3 } \left(\sqrt{6}-\frac{6 \sqrt{t_3}}{\sqrt{\eta\, R^2+6\, t_3}}\right).
   \end{equation}
   \subsubsection*{Comparison to the result from holography}
   The entanglement entropy from holography is given by eq. \eqref{holoEEd3} and may be expressed in terms of field theory quantities using eq. \eqref{gravityfieldtheory} (with $6\,\lambda_{3}=\ell_p^{2}\,\sqrt{6\,t_3}, \,L=\sqrt{6\,t_3}$)
   \begin{equation}
      S_{EE}= \frac{4\,\pi^2\,t_3}{\lambda_3}\,\eta\,\left(-\sqrt{6\,t_3}+\sqrt{6\,t_3+\eta\,R^2}\right).
   \end{equation}
   We see that the field theory calculation and the results from holography match up to a scheme dependent area term $\sim -4\,t_3\pi^2\,R/\lambda_3$. We obtain the exact same contribution from the Wald entropy associated with the counterterms given in eq. \eqref{wald1} which yields (in field theory variables) exactly  $\sim -4\,t_3\pi^2\,R/\lambda_3$. The entanglement entropies on both sides match, if the contributions of the counterterms -- which have been added to the field theory side -- are also taken into account in the gravitational theory.
   Similar to the literature, we may compare scheme independent quantities aka the renormalized entanglement entropy.
From the entanglement entropy, we immediately obtain the renormalized entanglement entropy by using eq. \eqref{runningC}
    \begin{equation}
      S_{\text{R,EE}}= \frac{4\, \pi^2\,\eta\,  t_3^{3/2}}{\lambda _3} \left(\sqrt{6}-\frac{6\,\sqrt{t_3}}{\sqrt{\eta  R^2+6\,t_3}}\right).
    \end{equation}
    We see that the results from holography and field theory perfectly match one another for the negative sign of the square root $\eta=1,\bm{s}=-1$ (AdS) and $\eta=-1,\bm{s}=-1$ (dS), respectively. 
\subsection{$d=4$}
In $d=4$ we have using eq. \eqref{omegaeq}
\begin{equation}
\omega_4=\frac{3\,\left(R^2+8\,t_6+R\,\bm{s}\sqrt{\eta\,R^2+16\,t_4}\right)}{8\,R^2\,\lambda_4}.
\end{equation}
    We can compute the sphere partition function by integrating with respect to $R$, where we fix the integration constant by demanding that $\log Z_{S^d}(R=0)=0$
\begin{align}
    \log Z_{S^4}=&-\frac{\pi^2}{\lambda_4}\left(R\left(R^3+16\,R\,t_4+R^2\,\bm{s}\,\sqrt{\eta\,R^2+16\,t_4}+\eta\,8\,\bm{s}\,t_4\sqrt{\eta\,R^2+16\,t_4}\right)\right.\nonumber\\&\left.-128\,\eta\,\bm{s}\,t_4^2\,\arcsinhh\left(\frac{R}{4\,\sqrt{t_4}}\right)\right).
\end{align}
We obtain the entanglement entropy by using the replica trick \eqref{EEreplica}. This gives us
\begin{equation}
S_{4,\text{EE}}=\frac{8\pi^2\,t_4}{\lambda_4}\left(R\left(-R+\eta\,\sqrt{\eta\, R^2+16\,t_4}\right)-16\,\eta\,t_4\,\arcsin(h)\left(\frac{R}{4\,\sqrt{t_4}}\right)\right).\label{EEEE4}
\end{equation}
 In $d=4$ dimensions, the renormalized entanglement entropy follows from eq. \eqref{EEEE4} with eq. \eqref{runningC}
   \begin{equation}
S_{\text{R,EE}}= \frac{128\, \pi ^2\, R^3\,  t_4^2}{\lambda_4  \left(\eta\, R^2+16 t_4\right)^{3/2}}.
   \end{equation}
   \subsubsection*{Comparison to the result from holography}
  In holography, the entanglement entropy in $d=4$ is given by eq. \eqref{holoEEd4} which reads in field theory quantities by relating $8\,\lambda_4=\ell_p^{3}\,\sqrt{16\,t_4}, \,L=\sqrt{16\,t_4}$
  \begin{equation}
       S_\text{EE}=\frac{8\,\pi^2\,t_4}{\lambda_4}\,\eta\,\left(R\sqrt{\eta\,R^2+16\,t_4}-16\,t_4\,\arcsinhh\frac{R}{4\,\sqrt{t_4}}\right).
   \end{equation}
   Again, this matches exactly our field theory computation up to a scheme dependent area term $-8\pi^2 R^2t_4/\lambda_4$ for the negative sign of the the square root. The area term with the negative sign comes from adding counterterms to our action. If we also consider the contributions of the counterterms in the gravitational theory eq. \eqref{wald2}, we see that we observe the exact same term there and thus the results of both sides match.
From the holographic entanglement entropy, we may derive the scheme independent entanglement entropy using eq. \eqref{runningC}
\begin{equation}
  S_{\text{R,EE}}=\frac{128 \pi ^2 R^3 t_4^2}{\lambda _4 \left(\eta\,R^2+16 t_4\right){}^{3/2}}
\end{equation}
We see that the renormalized entanglement entropies from field theory and holography in $d=4$ match perfectly for $\eta=1,\bm{s}=-1$ (AdS) and $\eta=-1,\bm{s}=-1$ (dS).

\subsection{$d=5$}
In $d=5$, the counterterm proportional to $c_d^{(3)}$ contributes for the first time. We find $\omega_5$ from eq. \eqref{omegaeq}
\begin{equation}
    \omega_5=\frac{30\,R^2\,t_5-225\,t_5^2\,R^3\left(R+\bm{s}\sqrt{\eta R^2+30\,t_5}\right)}{5\,R^4\,\lambda_5}.
\end{equation}
With $\omega_5$, we may compute the partition function by integrating eq. \eqref{partf} with respect to $R$ which results in
\begin{align}
\log Z_{S^5}=&-\frac{\pi^3}{5\,\lambda}\,\left(20\,\eta\,R^2\,\bm{s}\,t_5\sqrt{\eta\,R^2+30\,t_5}+1200\,\bm{s}\,t_5^2\left(\sqrt{30\,t_5}-\sqrt{\eta\,R^2+30\,t_5}\right)\right.\nonumber\\ 
& \left. +\left(2\,R^5+50\,R^3\,t_5-1125\,R\,t_5^2+2\,R^4\,\bm{s}\,\sqrt{\eta\,R^2+30\,t_5}\right)\right),
\end{align}
where we fixed the integration constant so that $\log Z_{S^5}(R=0)=0$. The entanglement entropy follows from the partition function using eq. \eqref{EEreplica}
\begin{align}
  S_{5,\text{EE}}=\frac{4\pi^3\,t_5}{\lambda_5}  \left(-R^3\,+45\,R\,t_5+\eta\,R^2\,\sqrt{\eta\,R^2+30\,t_5}+60\,t_5\left(\sqrt{30\,t_5}-\sqrt{\eta\,R^2+30\,t_5}\right)\right).
\end{align}
 In $d=5$ dimensions, we may compute the renormalized entanglement entropy using \eqref{runningC}
   \begin{equation}
S_{\text{R,EE}}=-\frac{240\, \pi ^3\, t_5^{5/2}\left(900\,t_5^{3/2}\!\!-30t_5\,\sqrt{900\,t_5+30\,\eta\,R^2}+\eta\,R^2\!\left(45\,\sqrt{t_5}-\sqrt{900t_5+\eta\,R^2}\right)\right)}{ \lambda_5  \left(\eta\, R^2+30 t_5\right)^{3/2}}.
   \end{equation}
 
 \subsubsection*{Comparison to the result from holography}
 To compare with the field theory result, we rewrite the result from holography \eqref{holoEEd5} with the dictionary $10\,\lambda_d=\ell_p^{4}\,\sqrt{30\,t_5}, \,L=\sqrt{30\,t_5}$ in field theory quantities
\begin{equation}
S_\text{EE}=\frac{4\,\pi^3\,t_5}{\lambda_5}\,\left(2\,(30\,t_5)^{3/2}+(\eta\,R^2-60\,t_5)\,\sqrt{30\,t_5+\eta\,R^2}\right).
\end{equation}
We see that the result from holography matches the calculation from field theory up to the scheme dependent terms $\sim 4\pi^3\,t_5/\lambda_5\,(-R^3\,+45\,R\,t_5)$. However, taking the contributions of the counterterms in the gravitational theory into account, we see find the exact same contribution to the entanglement entropy as observed in eq. \eqref{wald3}. The results of both sides hence match. For the sake of completeness, we calculate the renormalized entanglement entropy by using eq. \eqref{runningC}
   \begin{equation}
     S_{\text{R,EE}}=\frac{240 \pi ^3 t_5^2}{\lambda _5 \left(\eta\,  R^2+30\, t_5\right){}^{3/2}} \left( \sqrt{30\,t_5}\,\sqrt{\eta\,  R^2+30\, t_5}\left(30\, t_5 + \eta\,  R^2 \right)-45 \eta  R^2 t_5-900 t_5^2\right),
   \end{equation}
we see that the scheme dependent terms vanish and the results from field theory and holography agree perfectly for $\eta=1,\bm{s}=-1$ (AdS) and $\eta=-1,\bm{s}=-1$ (dS). 

\subsection{$d=6$}
In $d=6$, $\omega_6$ is given by eq. \eqref{omegaeq}
\begin{equation}
    \omega_6=\frac{5\left(R^4+24\,R^2\,t_6-288\,t_6^2+R^3\,\bm{s}\,\sqrt{\eta\,R^2+48\,t_6}\right)}{12\,R^4\,\lambda_6}.\label{o6}
\end{equation}
The partition function follows by inserting eq. \eqref{o6} into eq. \eqref{partf} and integrating with respect to $R$  
\begin{align}
\log Z_{S^6}=&-\frac{4 \pi ^3}{9 \lambda_6 } \left(R \left(R^5+36\,R^3\,t_6+R^4\,\bm{s}\sqrt{\eta\,R^2+48\,t_6}-864\,\bm{s}\,t_6^2\,\sqrt{-\eta\,R^2+48\,t_6}\right)\right. \nonumber\\
&\left.-864\,R\,t_6^2+\eta\,12\,R^2\,\bm{s}\,t_6\,\sqrt{\eta\,R^2+48\,t_6}+41472\,\bm{s}\,t_6^3\, \arcsinhh \left(\frac{R}{4 \sqrt{3\,t_6} }\right)\right),
\end{align}
where we chose the integration constant so that $\log Z_{S^6}(R=0)=0$. The entanglement follows from the partition function by eq. \eqref{EEreplica}
\begin{align}
    S_{EE}=&-\frac{16\,\pi^3\,t_6}{3\lambda_6}\left(R\left(R^3-48\,R\,t_6-\eta\,R^2\sqrt{\eta\,R^2+48\,t_6}+72\,t_6\,\sqrt{\eta\,R^2+48\,t_6}\right)\right.\nonumber\\
    &\left.-3456\,t_6^2\,\arcsinhh\left(\frac{R}{4\,\sqrt{3\,t_6}}\right)\right).
\end{align}
In $d=6$ dimensions, the renormalized entanglement entropy reads (using eq. \eqref{runningC})
   \begin{equation}
 S_{\text{R,EE}}=\frac{18432\, \pi^3\, R^5\, t_6^3}{\lambda  \left(\eta\,R^2+48 t_6\right)^{5/2}}.
   \end{equation}
   
  \subsubsection*{Comparison to the result from holography}
  The entanglement entropy from holography \eqref{holoEEd6} reads in $d=6$ in field theory quantities $12\,\lambda_6=\ell_p^{5}\,\sqrt{48\,t_6}$ and $L=\sqrt{48\,t_6}$
  \begin{equation}
      S_\text{EE}=\frac{8\,\pi^3\,t_6}{3\,\lambda_6}\,\left(R\,\sqrt{48\,t_6+\eta\,R^2}\,(2\,\eta\,R^2-144\,t_6)+6912 t_6^2\,\arcsinhh\left(\frac {R}{4\, \sqrt{3\,t_6}}\right)\right).
  \end{equation}
       The entanglement entropy from field theory matches the result from holography up to the usual area term $\sim16\,\pi^3\,t_6\,R^4/(3\lambda_6)$ and a scheme dependent term $\sim 256\, \pi^3\, R^2 t_6^2/\lambda_6$.
  The exact same terms arise in the gravitational theory too if we also take the counterterms into account there. The contributions are calculated in eq. \eqref{wald4} and match exactly the missing terms. We thus conclude that the entanglement entropies of both sides match. For comparison with similar results in the literature, we are looking at the renormalized entanglement entropy in $d=6$. We find a perfect match between field theory and the result from holography given by
  \begin{equation}
  S_{\text{R,EE}}=   \frac{18432\, \pi^3\, R^5\, t_6^3}{\lambda _6\, \left(\eta\,R^2+48 \,t_6\right){}^{5/2}},
  \end{equation}
 for $\eta=1,\bm{s}=-1$ (AdS) and $\eta=-1,\bm{s}=-1$ (dS).
\bibliographystyle{JHEP}
\bibliography{dsdsee}
\end{document}